\documentclass[floatfix,twocolumn,showpacs,superscriptaddress,nofootinbib,prd,aps,tightenlines]{revtex4-1}
\usepackage{graphicx}
\usepackage{psfrag}
\usepackage{dcolumn}
\usepackage{bm}
\usepackage{amsmath}
\usepackage{bm}
\usepackage{appendix}

\newcommand{\bea}{\begin{eqnarray}}
\newcommand{\eea}{\end{eqnarray}}
\newcommand{\beq}{\begin{equation}}
\newcommand{\eeq}{\end{equation}}

\usepackage{color}

\begin{document}

\title{Eccentricity estimation from initial data for Numerical Relativity Simulations}

\author{Alessandro Ciarfella}
\affiliation{Center for Computational Relativity and Gravitation,
School of Mathematical Sciences,
Rochester Institute of Technology, 85 Lomb Memorial Drive, Rochester,
New York 14623, USA}
  \author{James Healy}
\affiliation{Center for Computational Relativity and Gravitation,
School of Mathematical Sciences,
Rochester Institute of Technology, 85 Lomb Memorial Drive, Rochester,
New York 14623, USA}
  \author{Carlos O. Lousto}
\affiliation{Center for Computational Relativity and Gravitation,
School of Mathematical Sciences,
Rochester Institute of Technology, 85 Lomb Memorial Drive, Rochester,
New York 14623, USA}
  \author{Hiroyuki Nakano}
\affiliation{Faculty of Law, Ryukoku University, Kyoto 612-8577, Japan}

\date{\today}

\begin{abstract}
We describe and study an instantaneous definition of eccentricity to
be applied at the initial moment of full numerical simulations of
binary black holes. The method consists of evaluating the eccentricity
at the moment of maximum separation of the binary.  We estimate it
using up to third post-Newtonian (3PN) order, and compare these
results with those of evolving (conservative) 3PN equations of motion
for a full orbit and compute the eccentricity $e_r$ from the radial
turning points, finding excellent agreement. We next include terms
with spins up to 3.5PN, and then compare this method with the
corresponding estimates of the eccentricity $e_r^{NR}$ during full
numerical evolutions of spinning binary black holes, characterized
invariantly by a fractional factor $0\leq f\leq1$ of the initial
tangential momenta. It is found that our initial instantaneous
definition is a very useful tool to predict and characterize even
highly eccentric full numerical simulations.
\end{abstract}

\pacs{04.25.dg, 04.25.Nx, 04.30.Db, 04.70.Bw}\maketitle

\section{Introduction}\label{sec:Intro}

The concept of eccentricity is uniquely defined in Newtonian gravity.
An extension to General Relativity is not strictly uniquely or even
well defined, but we have found it useful to have a relationship to
estimate an instantaneous eccentricity, $e$, defined at initial data.

\begin{figure}[h!]
\hfill{\includegraphics[width=.4\textwidth]{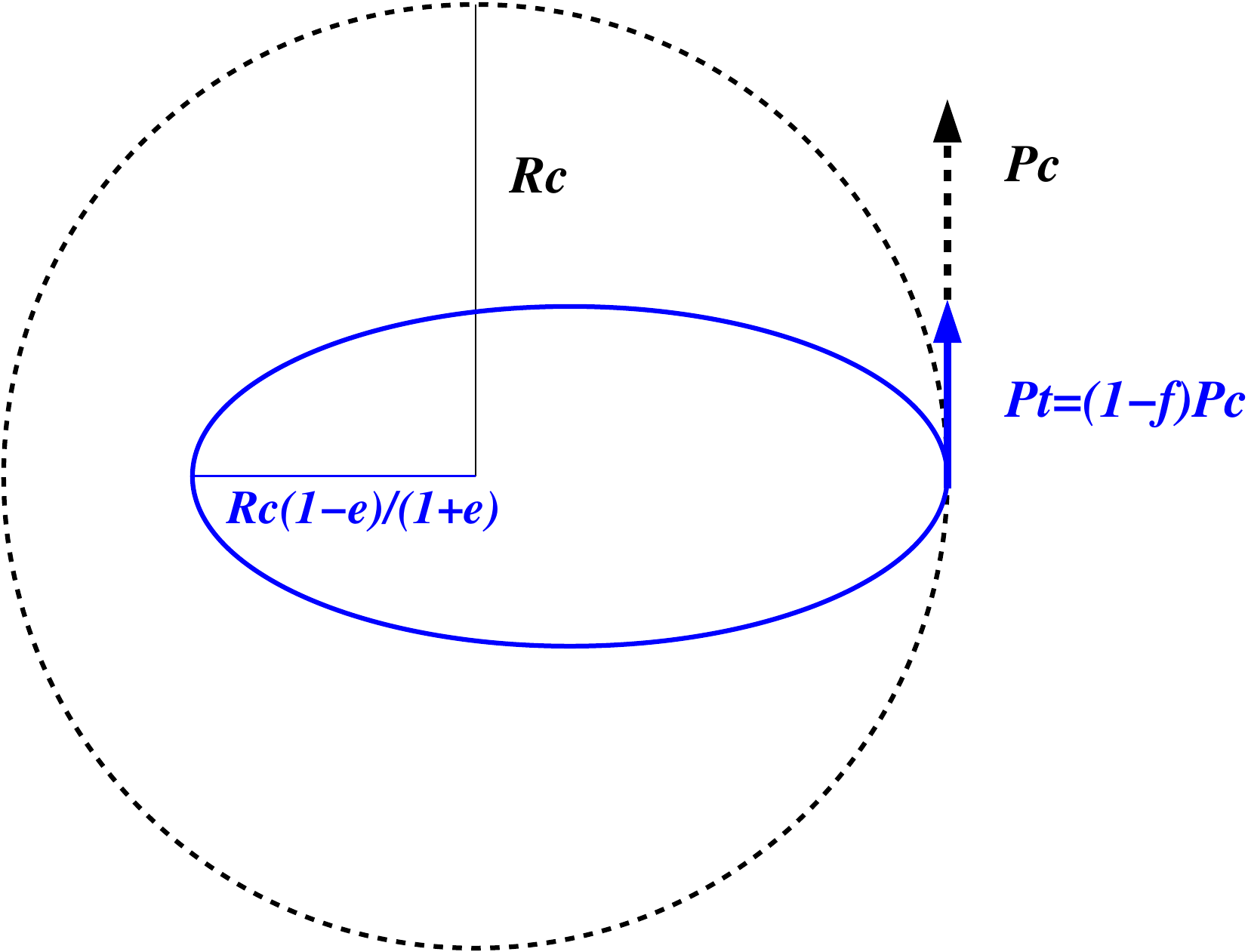}}
\caption{Schematic of the initial momentum choice to describe
  eccentric orbits in Numerical Relativity. $R_c$ and $P_c$ denote the
  initial separation and tangential momentum for a quasicircular
  binary. Then, $P_t=(1-f) P_c$ where $f$ is a fractional parameter,
  $0<f<1$, is the initial tangential momentum for an eccentric binary
  with an instantaneous eccentricity, $e$.}
\label{fig:ID}
\end{figure}

To compute the numerical initial data, we use the puncture
approach~\cite{Brandt97b} along with the {\sc TwoPunctures}
~\cite{Ansorg:2004ds} code. For each eccentric family of simulations,
we first determine the initial separation, $R_c$, and tangential
quasicircular momentum, $P_c$ using the results of
\cite{Healy:2017zqj}. To increase the eccentricity of the system while
keeping the initial data at an apocenter, the initial tangential
momentum, $P_t$, is modified by a fractional parameter, $0\leq
f\leq1$, such that $P_t=(1-f) P_c$. See Fig.~\ref{fig:ID} for an
schematic representation.

This method was applied to the estimates of templates of the
LIGO-Virgo detection GW190521 \cite{GW190521discovery} in
Ref.~\cite{Gayathri:2020coq} and of the 824 simulations included in
the latest (4th release \cite{Healy:2022wdn}) RIT catalog of binary
black hole simulations.  In Refs.~\cite{Gayathri:2020coq} and
\cite{Healy:2022wdn}, the initial eccentricity was then approximately
evaluated by the Newtonian relationship $e = 2f-f^2$. In this paper,
we extend this definition to higher post-Newtonian (PN) orders to
improve the identification of highly eccentric simulations and test it
against full numerical evolutions.

\section{Method}

The idea of this method applied to PN expansions is to evaluate the
conserved Hamiltonian at the two radial turning points of a binary,
$r_{\pm}$, to evaluate $j$, the conserved angular momentum at those
points and relate the eccentric and circular values at the apastron,
$r_+$, by a factor $(1-f)$ as displayed in Fig.~\ref{fig:ID}, for the
full numerical case.

\subsection{Nonspinning case}

Let us begin with the nonspinning case, for which we can write the
reduced Hamiltonian $\mathcal{H} = H/\mu$ with $\mu = m_1m_2/(m_1+m_2)$,
\begin{eqnarray}
    \mathcal{H}(\textbf{r},\hat{\textbf{p}}) &=& \mathcal{H}_0(\textbf{r},\hat{\textbf{p}}) + \frac{1}{c^2}\mathcal{H}_1(\textbf{r},\hat{\textbf{p}}) + \frac{1}{c^4}\mathcal{H}_2(\textbf{r},\hat{\textbf{p}}) 
    \cr && + \frac{1}{c^6}\mathcal{H}_3(\textbf{r},\hat{\textbf{p}}) \,,
\end{eqnarray}
where explicit expressions for the reduced 3PN Hamiltonian are given 
in Ref.~\cite{Memmesheimer:2004cv} (see also Appendix~\ref{app:PNH})
with $\textbf{r} = \textbf{R}/(G M)$ and $\hat{\textbf{p}} = \textbf{P}/\mu$ where 
$\textbf{R}$ is the relative separation vector, $M=m_1+m_2$,
and $\textbf{P}$ is the linear momentum. 
Writing this Hamiltonian in polar coordinates, we see that it does not depend on the coordinate $\phi$, and so $\hat{p}_\phi$ is a conserved quantity and the motion will happen only on a plane and so $\hat{\textbf{p}} = (\hat{p}_r,\,\hat{p}_\phi/r,\, 0)$. 
Now $\hat{p}_r$ vanishes at the turning points $r_+$ and $r_-$, and we can write
\begin{equation}
    \mathcal{H}(\textbf{r}_\pm,\hat{\textbf{p}}) = \mathcal{H}(r_\pm,\hat{p}_\phi) = \mathcal{H}_(r_\pm,j) \,,
\end{equation}
where $j = \hat{p}_\phi$ is constant along the orbit.

We now define the eccentricity measure $e_r$ as
\begin{equation}\label{er}
  e_r 
  = \frac{r_+ - r_-}{r_+ + r_-} \,.
\end{equation}
Therefore, $r_-$ is given by
\begin{equation}
	r_- = r_+ \frac{1-e_r}{1+e_r} \,,
\end{equation}
(see Fig.~\ref{fig:ID}).

To simplify more the computation, we scale again the Hamiltonian,
the momentum, $\hat{\textbf{p}}$, and the $r$-coordinate as
\begin{equation}\label{eq:norm}
    \tilde{\mathcal{H}} = r_+ \mathcal{H} \,,\quad 
    \tilde{\textbf{p}} = \sqrt{r_+}\,\hat{\textbf{p}} \,,\quad 
    \tilde{\textbf{r}} = \frac{\textbf{r}}{r_+} \,.
\end{equation}
This allows us to rewrite the Hamiltonian as 
\begin{eqnarray}
  \tilde{\mathcal{H}}(\tilde{\textbf{r}},\tilde{\textbf{p}}) =&& \tilde{\mathcal{H}}_0(\tilde{\textbf{r}},\tilde{\textbf{p}}) + \alpha \tilde{\mathcal{H}}_1(\tilde{\textbf{r}},\tilde{\textbf{p}}) + \alpha^2\tilde{\mathcal{H}}_2(\tilde{\textbf{r}},\tilde{\textbf{p}}) \cr
  && +\alpha^3\tilde{\mathcal{H}}_3(\tilde{\textbf{r}},\tilde{\textbf{p}}) \,,
\end{eqnarray}
where $\alpha = 1/(c^2 r_+)$. 

The advantage of this rescaling is that in this way we explicitly
remove the value of $r_+$ from our problem.  This appears only in the
expression for $\alpha$. In particular, we have (in polar coordinates)
\begin{equation}\label{tilder}
    \tilde{r}_+ = 1 \,,\quad \tilde{r}_- = \frac{1-e_r}{1+e_r} \,.
\end{equation}

Now since the Hamiltonian is conserved along the orbit, we must have
\begin{equation}\label{H_conv}
    \tilde{\mathcal{H}}(\tilde{r}_+,\tilde{j}) - \tilde{\mathcal{H}}(\tilde{r}_-,\tilde{j})=0 \,,
\end{equation}
where 
\begin{equation}
    \tilde{j} = \frac{j}{\sqrt{r_+}} \,.
\end{equation}

Using Eq.~(\ref{tilder}), Eq.~(\ref{H_conv}), and specifying values for $\alpha$ and $\eta$, we have an expression for $\tilde{j}$ in terms of $e_r$. 
Finally, introducing a momentum suppression factor $f$ as
\begin{equation}
    \tilde{j}(e_r) = (1-f) \tilde{j}_C \,,
\end{equation}
where $\tilde{j}(0) = \tilde{j}_C$ for the circular orbit, we obtain
\begin{equation}\label{eq:f}
	f(e_r) = 1 - \frac{\tilde{j}(e_r)}{\tilde{j}_C} \,.
\end{equation}

This final expression provides us with the desired relationship to
evaluate $f(e_r)$ and to invert (numerically) for any specific set of
initial parameters of a binary black hole simulation and obtain the
estimated $e_r$.

\subsection{Spinning case}

For the spinning case, we can apply the same method. Let us consider two
orbiting, nonprecessing, black holes with spins $\textbf{S}_1$ and
$\textbf{S}_2$ in the direction of the orbital angular momentum. The
Hamiltonian becomes \cite{Tessmer:2010hp,Tessmer:2012xr} (we
restore here the explicit dependence with the speed of light $c$ to
better display PN orders),
\begin{align}
    \begin{split}
      \mathcal{H}&(\textbf{r},\textbf{p},\textbf{S}_1,\textbf{S}_2) =\\
      & \mathcal{H}_0(\textbf{r},\textbf{p}) +  \frac{1}{c^2}\mathcal{H}_1(\textbf{r},\textbf{p}) + \frac{1}{c^4}\mathcal{H}_2(\textbf{r},\textbf{p}) + \frac{1}{c^6}\mathcal{H}_3(\textbf{r},\textbf{p}) \\
    &+\frac{\delta}{c^2}\mathcal{H}_{SO}^{LO}(\textbf{r},\textbf{p},\textbf{S}_1,\textbf{S}_2) +
    \frac{\delta}{c^4}\mathcal{H}_{SO}^{NLO}(\textbf{r},\textbf{p}, \textbf{S}_1,\textbf{S}_2) \\
   &+\frac{\delta^2}{c^2}\mathcal{H}_{S1S2}^{LO}(\textbf{r},\textbf{p},\textbf{S}_1,\textbf{S}_2) +
    \frac{\delta^2}{c^2}\mathcal{H}_{S^2}^{LO}(\textbf{r},\textbf{p},\textbf{S}_1,\textbf{S}_2) \\
    &+\frac{\delta^2}{c^4}\mathcal{H}_{S1S2}^{NLO}(\textbf{r},\textbf{p},\textbf{S}_1,\textbf{S}_2) +
    \frac{\delta^2}{c^4}\mathcal{H}_{S^2}^{NLO}(\textbf{r},\textbf{p},\textbf{S}_1,\textbf{S}_2) \\
    &+\frac{\delta}{c^6}\mathcal{H}_{SO}^{NNLO}(\textbf{r},\textbf{p},\textbf{S}_1,\textbf{S}_2) \, +
     \frac{\delta^3}{c^4}\mathcal{H}_{S^3}^{LO}(\textbf{r},\textbf{p},\textbf{S}_1,\textbf{S}_2) \,.
    \end{split}
\end{align}
where $\delta$ is a dimensionless factor that keeps track of the spin order 
of the term considered (see also Appendix~\ref{app:PNH}).
In this case, we define
\begin{equation}
    \tilde{\textbf{S}}_a = \frac{\textbf{S}_a}{\sqrt{r_+}} = \sqrt{\alpha} \bm{\chi}_a \,,\quad (a=1,\,2) \,,
\end{equation}
where in the last equality we introduced the dimensionless quantity
$\bm{\chi}_a$ as 
\begin{equation}
    \bm{\chi}_a = \frac{\hat{\textbf{S}}_a}{m_a^2}\,,\quad (a = 1,\,2) \,.
\end{equation}

Here, $\hat{\textbf{S}}_a$ are the actual spins with dimension (geometric units) 
$[\hat{\textbf{S}}] = [({\rm Mass})]^2$.
In terms of this new dimensionless variable we have the rescaled Hamiltonian as 
\begin{align}
    \begin{split}
      \tilde{\mathcal{H}}&(\tilde{\textbf{r}},\tilde{\textbf{p}},\bm{\chi}_1,\bm{\chi}_2) =\\
      & \tilde{\mathcal{H}}_0(\tilde{\textbf{r}},\tilde{\textbf{p}}) +  \alpha\tilde{\mathcal{H}}_1(\tilde{\textbf{r}},\tilde{\textbf{p}}) + \alpha^2\tilde{\mathcal{H}}_2(\tilde{\textbf{r}},\tilde{\textbf{p}}) + \alpha^3\tilde{\mathcal{H}}_3(\tilde{\textbf{r}},\tilde{\textbf{p}})\\
      &+
    \alpha^{3/2}\tilde{\mathcal{H}}_{SO}^{LO}(\tilde{\textbf{r}},\tilde{\textbf{p}},\bm{\chi}_1,\bm{\chi}_2) +
    \alpha^{5/2}\tilde{\mathcal{H}}_{SO}^{NLO}(\tilde{\textbf{r}},\tilde{\textbf{p}}, \bm{\chi}_1,\bm{\chi}_2) \\
    &+
    \alpha^2\tilde{\mathcal{H}}_{S1S2}^{LO}(\tilde{\textbf{r}},\tilde{\textbf{p}},\bm{\chi}_1,\bm{\chi}_2) +
    \alpha^2\tilde{\mathcal{H}}_{S^2}^{LO}(\tilde{\textbf{r}},\tilde{\textbf{p}},\bm{\chi}_1,\bm{\chi}_2) \\
    &+
    \alpha^3\tilde{\mathcal{H}}_{S1S2}^{NLO}(\tilde{\textbf{r}},\tilde{\textbf{p}},\bm{\chi}_1,\bm{\chi}_2) +
    \alpha^3\tilde{\mathcal{H}}_{S^2}^{NLO}(\tilde{\textbf{r}},\tilde{\textbf{p}},\bm{\chi}_1,\bm{\chi}_2) \, \\
    &+
    \alpha^{7/2}\tilde{\mathcal{H}}_{SO}^{NNLO}(\tilde{\textbf{r}},\tilde{\textbf{p}},\bm{\chi}_1,\bm{\chi}_2) +
    \alpha^{7/2}\tilde{\mathcal{H}}_{S^3}^{LO}(\tilde{\textbf{r}},\tilde{\textbf{p}},\bm{\chi}_1,\bm{\chi}_2) \,.
    \end{split}
\end{align}

We can now follow the same steps as indicated in
Eqs.~(\ref{H_conv})--(\ref{eq:f}) to obtain a relationship between the
fractional parameter $f$ by which the tangential circular momentum is
suppressed to generate eccentric orbits, and the eccentricity $e_r$,
defined through the periastron and apastron (see also
Appendix~\ref{sec:1PN}).

\section{Results}

In the applications below, we will assume for the sake of definiteness 
and comparisons with the simulations used for GW190521 in
Ref.~\cite{Gayathri:2020coq} an initial coordinate separation
of the holes of about $r\approx24.7M$, that we will use in the evaluation of
$\alpha$ above. This corresponds in the cases
studied in Ref.~\cite{Gayathri:2020coq} to 
an initial quasicircular reference frequency
of 10\,Hz for a $30\,$M$_\odot$ system, as evaluated by the techniques
described in Ref.~\cite{Healy:2017zqj}.  

\subsection{Fractional parameter $f(e_r)$ for given initial parameters} 

\begin{figure}[h!]
\centering
\includegraphics[width=.49\textwidth]{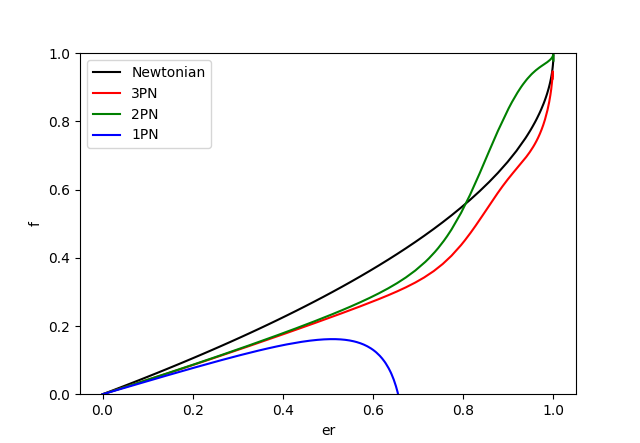}
\caption{Momentum suppression factor $f$ vs. eccentricity $e_r$ for
  nonspinning equal mass binaries, $\chi_i=0$ and $q=1$, at various PN
  orders.}
\label{fig:f_vs_er3}
\end{figure}

The result for nonspinning equal mass binaries, i.e., the mass ratio
$q = m_2/m_1 = 1$, at different successive
PN orders is shown in Fig.~\ref{fig:f_vs_er3}.
We plot here the factor $f$ by which we reduce the tangential linear momentum
of a quasicircular orbit versus the computed eccentricity $e_r$. This allow
us to read off the eccentricity associated to our initial data set up. We
can see the good agreement to all displayed PN orders at low eccentricities
$(e_r<0.4)$. At intermediate eccentricities the 1PN computation deviates
from the higher order trend for $e_r>0.4$, while the 2PN computation remains
consistent for $e_r<0.7$. The 3PN computation on the other hand converges towards
the Newtonian (0PN) curve for larger $e_r$.
We interpret this as the correct behavior since
for large $e_r$ the expected evolution of a binary is essentially a plunge that
tends to reduce the differences between PN orders.

\begin{figure}[h!]
\centering
\includegraphics[width=.49\textwidth]{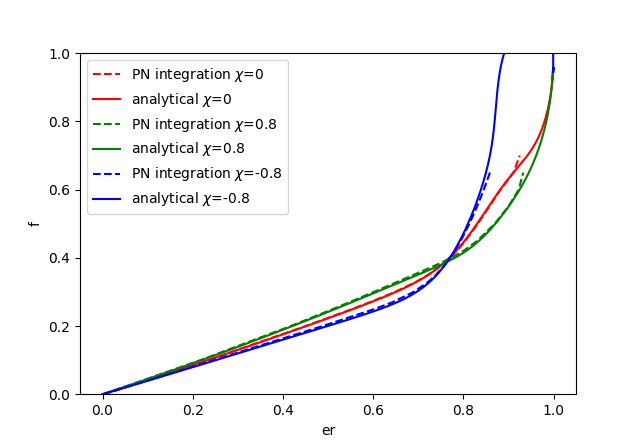}
\caption{Comparison of initial analytic vs. integration of the 3PN equations
  of motion: Eccentricity $e_r$ vs. momentum suppression factor $f$
for $q=1$ and different values of the spins.
}
\label{f_vs_er3}
\end{figure}

In the case of spinning holes, we will hence focus directly on the 3PN
computation and compare them with the nonspinning case. 
To better display the effects (and following comparisons with full numerics),
we will consider the cases when both equal mass black holes have spins
aligned $(\chi_i=+0.8)$ or anti-aligned $(\chi_i=-0.8)$ with the orbital
angular momentum (we checked that the $\chi_1=-\chi_2$ case gives a curve
very close to the nonspinning case). The results for this 3PN order
comparisons for the
various values of the spins are shown in Fig.~\ref{f_vs_er3}. 
We observe a close dependence of the three curves for small and intermediate
eccentricities, but for $e_r>0.75$ there is a reverse in their relative
behavior. While the case of aligned spins eventually merger for large
eccentricities (plunges) with the nonspinning holes, the antialigned
spins case shows a very different behavior. We will come back later
to this case with a 3.5PN computation.


\subsection{Comparison with numerical integrations of 3PN equations of motion} 

A first validation of our initial instantaneous eccentricity estimate
can be performed by comparing our analytical results with the
numerical integrations of the conservative 3PN equation of motion
\cite{Damour:2007nc,Buonanno:2005xu}, where we suppressed the 2.5PN
radiative terms. We integrate the orbital motion over the first orbit
and evaluate the eccentricity from the apastron and periastron
differences, $e_r=(r_+-r_-)/(r_++r_-)$. The comparisons for spinning
and nonspinning cases with $q=1$, and $\chi_i=0,\,\pm0.8$ are displayed in
Fig.~\ref{f_vs_er3}. The results show a notable agreement and
consistency between the integrated and initial estimates of the
eccentricity, for $e_r \leq 0.9$, at 3PN order.

\begin{figure}[h!]
\centering
\includegraphics[width=0.49\textwidth]{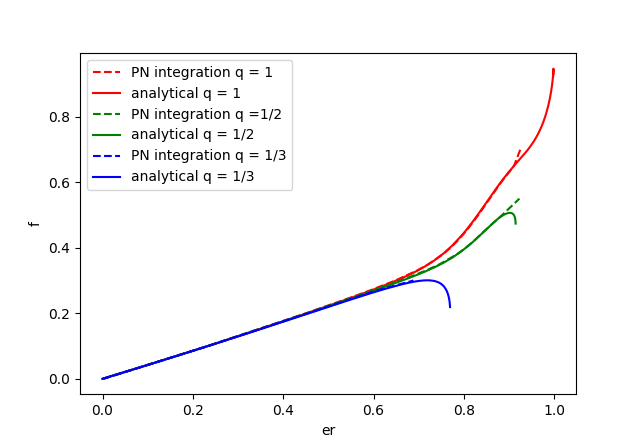}
\caption{Comparison of initial analytic vs. 3PN equation of motion
  integration of the momentum suppression factor $f$ vs.  eccentricity
  $e_r$ for nonspinning binaries, $\chi_i = 0$ and different values of
  the mass ratio $q$.}
\label{f_vs_er4exp}
\end{figure}

To verify the mass ratio dependence of our eccentricity
estimator as well, we compare our analytical results with 
numerical evolutions of the 3PN equations of
motion in Fig.~\ref{f_vs_er4exp} for mass ratios $q=1,\, 1/2$ and
$1/3$ (for nonspinning binaries). We observe again a notable agreement
in their corresponding regions or validity (as the 3PN approximation
reduces its validity to medium and small eccentricity as we deal with
smaller mass ratios). 

\subsection{Comparisons with full Numerical Relativity simulations}\label{sec:NR}

\begin{figure}[h!]
\centering
\includegraphics[width=0.49\textwidth]{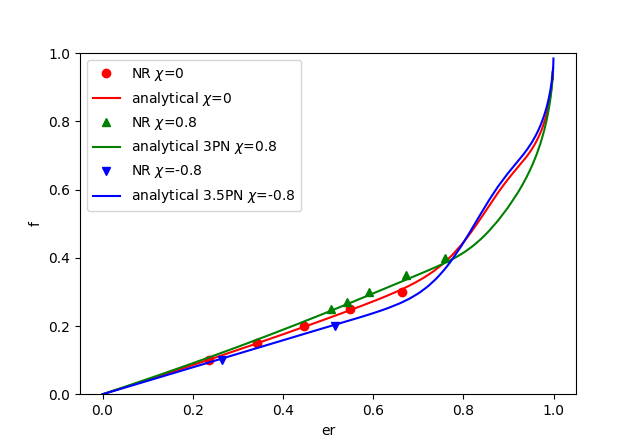}
\caption{Momentum suppression factor $f$ vs. eccentricity $e_r$ with PN 
estimates for various spins (continuous curves)
in comparison with the full numerical simulation measurements (dots).
}
\label{f_vs_eNR}
\end{figure}

We are now able to directly compare our initial eccentricity PN
estimates to actual full numerical simulations where it is possible to
evaluate the eccentricity via the turning points in the simulations.
We thus identify the numerical and PN (in ADMTT gauge) parameters,
$R_c=r_+$ and $P_c=P_\phi(e_r=0)/r_+$, and the values of $f$,
$\alpha$, $q$, and $S_1$ and $S_2$ for several simulations available
in the RIT waveforms catalog \cite{Healy:2022wdn} identified in Table
\ref{tab:ID}.  The results are displayed in Fig.~\ref{f_vs_eNR}. The
agreement for simulations in the range of low to middle eccentricities
is remarkable. We also include here the 3.5PN corrections to the
antialigned spins configurations to display an improved behavior all
the way up to $e_r\to1$, merging with the plunging behavior in the
cases of nonspinning and aligned spins.

\begin{table}
  \centering
 \caption{Eccentric simulations used in Fig.~\ref{f_vs_eNR} and their estimated eccentricities from its radial turning points, $e_r^{NR}$}
 \label{tab:ID}
 \begin{tabular}{llllll|l}
	 \hline
RIT Catalog No. & $R_c$ & $q$ & $\chi_1^z$ & $\chi_2^z$ & $f$ &\ $e_r^{NR}$\\ 
 	 \hline 
         RIT:eBBH:1282 & 24.64 & 1 & 0.0 & 0.0 & 0.10 & 0.2357\\ 
         RIT:eBBH:1283 & 24.64 & 1 & 0.0 & 0.0 & 0.15 & 0.3416\\
         RIT:eBBH:1285 & 24.64 & 1 & 0.0 & 0.0 & 0.20 & 0.4459\\
         RIT:eBBH:1293 & 24.64 & 1 & 0.0 & 0.0 & 0.25 & 0.5488\\
         RIT:eBBH:1303 & 24.64 & 1 & 0.0 & 0.0 & 0.30 & 0.6646\\
         RIT:eBBH:1807 & 24.56 & 1 & 0.8 & 0.8 & 0.25 & 0.5064\\
         RIT:eBBH:1808 & 24.56 & 1 & 0.8 & 0.8 & 0.27 & 0.5410\\
         RIT:eBBH:1809 & 24.56 & 1 & 0.8 & 0.8 & 0.30 & 0.5915\\
         RIT:eBBH:1811 & 24.56 & 1 & 0.8 & 0.8 & 0.35 & 0.6735\\
         RIT:eBBH:1813 & 24.56 & 1 & 0.8 & 0.8 & 0.40 & 0.7587\\
         RIT:eBBH:1763 & 24.75 & 1 & -0.8 & -0.8 & 0.10 & 0.2644 \\
         RIT:eBBH:1764 & 24.75 & 1 & -0.8 & -0.8 & 0.20 & 0.5143\\
         \hline 
\end{tabular} 
\end{table} 

\begin{figure}[h!]
\centering
\includegraphics[width=0.49\textwidth]{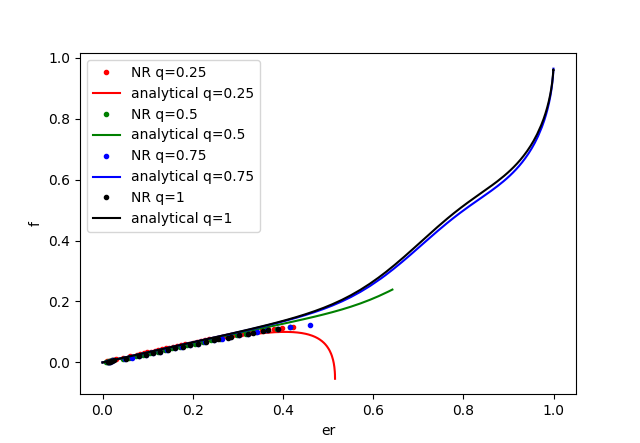}
\caption{Momentum suppression factor $f$ vs.  eccentricity $e_r$ for
  various mass ratio nonspinning binaries using 3PN estimates
  (continuous curves) at $r=11.35M$, in comparison with the
  corresponding full numerical simulations evaluations (dots).  }
\label{f_vs_eNR11}
\end{figure}

In the RIT catalog \cite{Healy:2022wdn}, we have another family of
eccentric simulations (for nonspinning and different mass ratios
$q=1,\,3/4,\,1/2$ and $1/4$), starting at much closer initial separations,
$r\approx11.35M$, that we can use to compare to our PN eccentric
estimations. These separations are roughly half the ones we considered
so far, and are at the limit of applicability of PN expansions.  The
results of these estimates are displayed in Fig.~\ref{f_vs_eNR11}.  It
is also difficult to compute the $e_r$ from the full numerical
evolutions for large eccentricities since the trajectories are highly
inspiral or merge before we can complete a meaningful orbit to extract
$r_+$ and $r_-$. Yet, the estimates are very good for the expected
range of validity of the PN expansions for small mass ratios (here
$e_r<0.5$).

We conclude that our eccentricity estimates provide an accurate description
of the initial binary black holes eccentric properties and can be directly
applied to the eccentric simulations in the 4th RIT catalog \cite{Healy:2022wdn}.

\section{Conclusions and Discussion}\label{sec:Discussion}

We have defined eccentric binary black hole simulations invariantly in
terms of fractional, $f=1-P_t(e_r)/P_c$, tangential linear momenta to
the circular one. We have found that the PN analytic estimates of the
initial eccentricity of these full numerical simulations are an
accurate and practical tool to predict and assess the first orbit
eccentricity of full numerical simulations, allowing, for instance,
precise design of new runs for parameter coverage or targeted studies.
For low and medium eccentricities $e_r<0.5$, and separated
enough binaries, even 2PN estimates are accurate. For higher
eccentricities and highly spinning (particularly for both antialigned)
binary black holes, we require 3PN, 3.5PN or even eventually 4PN
estimates at closer initial separations.
Our formalism can be also applied to generic orientations of the
spins by use of the concept of spherical orbits~\cite{Buonanno:2010yk}
to compute the turning points $r_\pm$.

Here, we have suppressed the tangential momentum with respect to the
quasicircular one by a $(1-f)$ factor,
with $0\leq f\leq1$. But if we allow for $f<0$, we would actually
increase the tangential momentum, leading to an elliptic orbit, but starting
at the periastron $(r_-)$ instead of the apastron $(r_+)$. This can
be achieved by reversing the sign of $e$ in our equations. For instance,
for the Newtonian case we would have $e_r=-2f+f^2$ for $F_p< f \leq0$,
where in this case $F_p=1-\sqrt{2}=-0.4142...$ would lead to a parabola.
For values more negative
than this $F_p$, i.e., $f<F_p$, we would generate a hyperbolic orbit.

The estimates which we have developed can now be directly applied to
the 824 eccentric simulations in the 4th RIT catalog
\cite{Healy:2022wdn}.  Our formulas should still provide good
estimates for well separated precessing binaries with small radial
momentum components by use of the projected spins along the initial
orbital angular momentum as variables. This is the case for all our
simulations in Ref.~\cite{Healy:2022wdn}, and in particular we can now
reassess the best eccentricity estimate of the gravitational waves
event GW190521 \cite{Gayathri:2020coq}. In that paper, we assessed the
eccentricity of the optimal full numerical simulation with the
Newtonian estimate to be $e_N=0.69$ (with $q=1$,
$\chi_1^z=\chi_2^z=0.27$, $R_c=24.7M$, and $f=0.44$). We can now
recompute the eccentricity using our 3.5PN estimator and find
$e_{PN}=0.80$, which highlights again the potentially interesting
astrophysical scenarios that might have lead to the merger of the two
black holes generating GW190521
\cite{LIGOScientific:2020ufj,Barrera:2022yfj}.

\begin{acknowledgments}

The authors gratefully acknowledge the National Science Foundation
(NSF) for financial support from Grant No.\ PHY-1912632.
Computational resources were also provided by the NewHorizons, BlueSky
Clusters, GreenPrairies, and WhiteLagoon at the Rochester Institute
of Technology, which were supported by NSF grants No.\ PH-0722703,
No.\ DMS-0820923, No.\ AST-1028087, No.\ PHY-1229173,
No.\ PH-1726215, and No.\ PH-2018420.  This work used the Extreme
Science and Engineering Discovery Environment (XSEDE) [allocation
  TG-PHY060027N], which is supported by NSF grant No.\ ACI-1548562 and
project PHY20007 Frontera, an NSF-funded petascale computing system at
the Texas Advanced Computing Center (TACC).
H.N. is supported by JSPS/MEXT KAKENHI Grant 
No.\ JP21K03582, No.\ JP21H01082, and No.\ JP17H06358.

\end{acknowledgments}



\appendix

\section{PN Hamiltonian}\label{app:PNH}

In this appendix we provide the explicit form of the Hamiltonian terms
(up to 2PN order) that we used throughout this paper. 
From Eq.~(7) in Ref.~\cite{Memmesheimer:2004cv}, 
we have the nonspinning components of the Hamiltonian,
\begin{equation}
\mathcal{H}_0(\textbf{r},\hat{\textbf{p}}) = \frac{\hat{\textbf{p}}^2}{2} - \frac{1}{r} \,,
\end{equation}
\begin{eqnarray}
\mathcal{H}_1(\textbf{r},\textbf{p}) &=& \frac{1}{8}(3\eta - 1)(\hat{\textbf{p}}^2)^2 - \frac{1}{2r}\left[(3+\eta)\hat{\textbf{p}}^2 + \eta(\textbf{n}\cdot{\hat{\textbf{p}}})^2\right]  
\cr && +\frac{1}{2r^2} \,,
\end{eqnarray}
\begin{align}
\begin{split}
&\mathcal{H}_2(\textbf{r},\textbf{p}) = \frac{1}{16}(1-5\eta+5\eta^2)(\hat{\textbf{p}}^2)^3 
\\ &+ \frac{1}{8r}\left[(5 - 20\eta - 3\eta^2)(\hat{\textbf{p}}^2)^2 - 2\eta^2(\textbf{n}\cdot{\hat{\textbf{p}}})^2\hat{\textbf{p}}^2-3\eta^2(\textbf{n}\cdot\hat{\textbf{p}})^4\right] \\
&+\frac{1}{2r^2}\left[(5+8\eta)\hat{\textbf{p}}^2 + 3\eta(\textbf{n}\cdot{\hat{\textbf{p}}})^2\right] -\frac{1}{4r^3}(1+3\eta) \,,
    \end{split}
\end{align}
where $\eta=m_1 m_2/(m_1+m_2)^2$.

The explicit expressions for the spin terms of the Hamiltonian are given in Eqs.~(13)--(16) of Ref.~\cite{Tessmer:2010hp} and Eqs.~(15)--(18) of Ref.~\cite{Tessmer:2012xr}. Here we write some of them in the notation used throughout this paper
\begin{align}
    \begin{split}
      \mathcal{H}_{SO}^{LO}(\textbf{r},\textbf{p}) =& \frac{1}{r^3}\left[\left(1-\frac{\eta}{2}+\sqrt{1-4\eta}\right)\left(\textbf{h}\cdot\textbf{S}_1\right)\right.\\
        &\left.+ \left(1-\frac{\eta}{2}-\sqrt{1-4\eta}\right)\left(\textbf{h}\cdot\textbf{S}_2\right)\right],
    \end{split}
\end{align}
\begin{align}
    \begin{split}
&\mathcal{H}_{S^2}^{LO}(\textbf{r},\textbf{p}) = \\
&\frac{\eta}{r^3}\bigg[\lambda_1 \left(-1+2\eta-\sqrt{1-4\eta}\right)\left(3\left(\textbf{n}_{12}\cdot\textbf{S}_1\right)^2 - \left(\textbf{S}_1\cdot\textbf{S}_1\right)\right)\\
        &+\lambda_2 \left(-1+2\eta+\sqrt{1-4\eta}\right)\left(3\left(\textbf{n}_{12}\cdot\textbf{S}_2\right)^2- \left(\textbf{S}_2\cdot\textbf{S}_2\right)\right)\bigg],
    \end{split}
\end{align}
\begin{align}
    \begin{split}
            \mathcal{H}_{S1S2}^{LO}(\textbf{r},\textbf{p}) =& \frac{\eta}{r^3}\left(3\left(\textbf{n}_{12}\cdot\textbf{S}_1\right)\left(\textbf{n}_{12}\cdot\textbf{S}_2\right) - \left(\textbf{S}_1\cdot\textbf{S}_2\right)\right),
    \end{split}
\end{align}
where for BHs $\lambda_1 = \lambda_2 = -1/2$, $\textbf{n}_{12} = \textbf{r}/|r|$ 
and $\textbf{h} = r\textbf{n}_{12}\times\hat{\textbf{p}}$.

\section{Scripts/Notebooks}\label{sec:script}

Here, we present a minimalistic script to compute the eccentricity from
the full numerical simulation parameters $q,\,R,\,\chi_1^z,\,\chi_2^z$ and $f$.
For the sake of simplicity we only include explicitly up to the 2PN
Hamiltonian terms, but in the results of the paper we computed up to
3.5PN terms. The script only allows for spins oriented along the z axis but it can be extended in order to include any orientation of the spins. 4PN local terms can be added in a straightforward way too, but the non local (see Ref.~\cite{Schafer:2018kuf}) terms are more difficult to include in our formalism.

\begin{figure*}[ht!]
\centering
\includegraphics[width=0.49\textwidth]{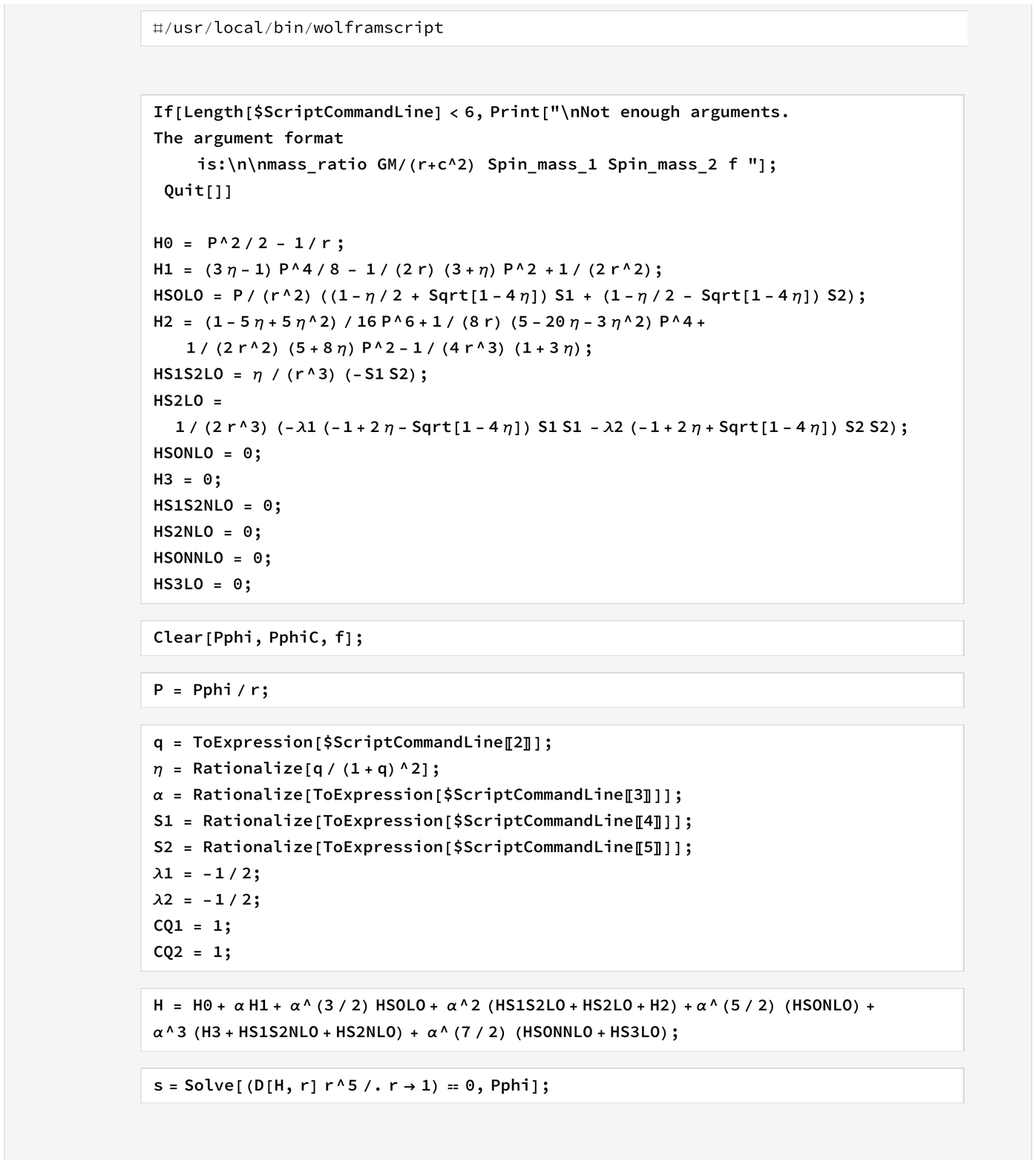}
\includegraphics[width=0.49\textwidth]{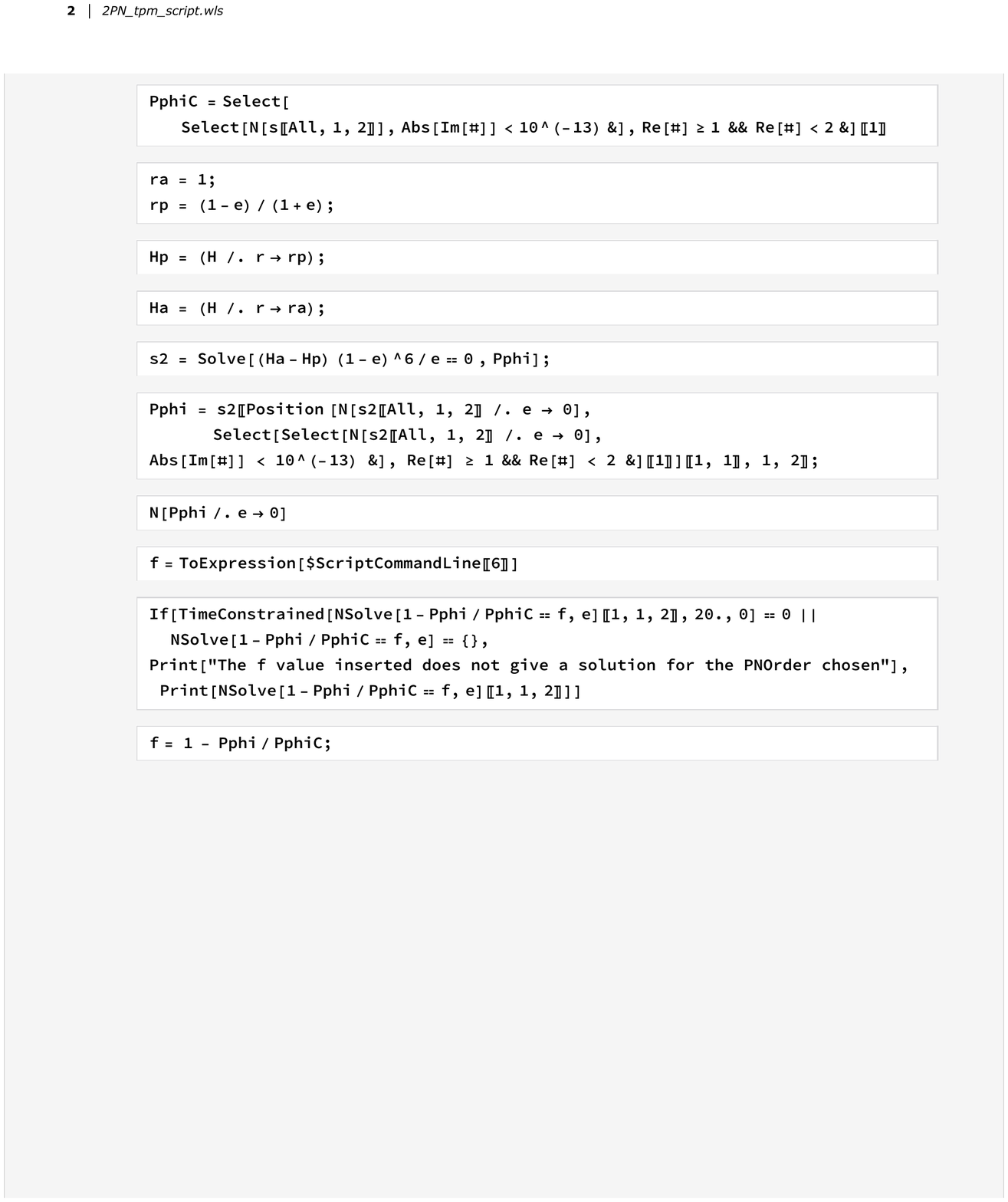}
\caption{Script to evaluate the eccentricity (explicit to 2PN order). 
}
\label{fig:script}
\end{figure*}



\section{Explicit analytic expressions for 1PN}\label{sec:1PN}

Here, we derive explicit analytic expressions for $f(e_r)$ at
a lower PN order expansions in the eccentricity, 
We hence consider the 1PN Hamiltonian,
\bea
H &=& \frac12\,{\frac {{P_{{\phi}}}^{2}}{{r}^{2}}}-\frac1r
\cr && +\alpha \left[\frac18
\,{\frac { \left( 3\,\eta-1 \right) {P_{{\phi}}}^{4}}{{r}^{4}}}-\frac12\,{
\frac { \left( 3+\eta \right) {P_{{\phi}}}^{2}}{{r}^{3}}}+\frac{1}{r^2} \right] \,. 
\eea

From equating the values of the Hamiltonian at the periastron and apastron, $r_+$,
\beq
H(r=r_+,P_{\phi}^2)=H(r=r_+(1-e_r)/(1+e_r),P_{\phi}^2) \,,
\eeq
picking up the right root of $P_{\phi}^2$, 
we find for our approximation of $f(e_r)=1-\sqrt{P_{\phi}^2/P_{\phi}^2(e_r=0)}$
\beq
f(e_r)=-\frac{\left[-\Delta+\frac{\alpha  (22-6 \eta )-4}{\Delta}-3 \alpha  (\eta +3)+4\right] e_r}{2 \left(\Delta+3 \alpha  (\eta +3)-2\right)} \,,
\eeq
where
\beq
\Delta=\sqrt{\alpha  \left(\alpha  \left(9 \eta ^2+30 \eta +89\right)+12 \eta -44\right)+4} \,.
\eeq
This expression is formally valid for $e_r<0.3$,
but it can be used as a first estimate up to intermediate eccentricities,
$e_r<0.6$ in the large separation regime $r_+>12M$, comparable masses,
$q>1/4$, and slowly spinning black holes $\chi_i<0.5$, as we verified
by direct comparisons with full 3PN expressions.

\bibliographystyle{apsrev4-1}
\bibliography{../../../Bibtex/references.bib}

\begin{thebibliography}{15}%
\makeatletter
\providecommand \@ifxundefined [1]{%
 \@ifx{#1\undefined}
}%
\providecommand \@ifnum [1]{%
 \ifnum #1\expandafter \@firstoftwo
 \else \expandafter \@secondoftwo
 \fi
}%
\providecommand \@ifx [1]{%
 \ifx #1\expandafter \@firstoftwo
 \else \expandafter \@secondoftwo
 \fi
}%
\providecommand \natexlab [1]{#1}%
\providecommand \enquote  [1]{``#1''}%
\providecommand \bibnamefont  [1]{#1}%
\providecommand \bibfnamefont [1]{#1}%
\providecommand \citenamefont [1]{#1}%
\providecommand \href@noop [0]{\@secondoftwo}%
\providecommand \href [0]{\begingroup \@sanitize@url \@href}%
\providecommand \@href[1]{\@@startlink{#1}\@@href}%
\providecommand \@@href[1]{\endgroup#1\@@endlink}%
\providecommand \@sanitize@url [0]{\catcode `\\12\catcode `\$12\catcode
  `\&12\catcode `\#12\catcode `\^12\catcode `\_12\catcode `\%12\relax}%
\providecommand \@@startlink[1]{}%
\providecommand \@@endlink[0]{}%
\providecommand \url  [0]{\begingroup\@sanitize@url \@url }%
\providecommand \@url [1]{\endgroup\@href {#1}{\urlprefix }}%
\providecommand \urlprefix  [0]{URL }%
\providecommand \Eprint [0]{\href }%
\providecommand \doibase [0]{http://dx.doi.org/}%
\providecommand \selectlanguage [0]{\@gobble}%
\providecommand \bibinfo  [0]{\@secondoftwo}%
\providecommand \bibfield  [0]{\@secondoftwo}%
\providecommand \translation [1]{[#1]}%
\providecommand \BibitemOpen [0]{}%
\providecommand \bibitemStop [0]{}%
\providecommand \bibitemNoStop [0]{.\EOS\space}%
\providecommand \EOS [0]{\spacefactor3000\relax}%
\providecommand \BibitemShut  [1]{\csname bibitem#1\endcsname}%
\let\auto@bib@innerbib\@empty
\bibitem [{\citenamefont {Brandt}\ and\ \citenamefont
  {Br{\"u}gmann}(1997)}]{Brandt97b}%
  \BibitemOpen
  \bibfield  {author} {\bibinfo {author} {\bibfnamefont {S.}~\bibnamefont
  {Brandt}}\ and\ \bibinfo {author} {\bibfnamefont {B.}~\bibnamefont
  {Br{\"u}gmann}},\ }\href@noop {} {\bibfield  {journal} {\bibinfo  {journal}
  {Phys. Rev. Lett.}\ }\textbf {\bibinfo {volume} {78}},\ \bibinfo {pages}
  {3606} (\bibinfo {year} {1997})},\ \Eprint
  {http://arxiv.org/abs/gr-qc/9703066} {gr-qc/9703066} \BibitemShut {NoStop}%
\bibitem [{\citenamefont {Ansorg}\ \emph {et~al.}(2004)\citenamefont {Ansorg},
  \citenamefont {Br\"ugmann},\ and\ \citenamefont {Tichy}}]{Ansorg:2004ds}%
  \BibitemOpen
  \bibfield  {author} {\bibinfo {author} {\bibfnamefont {M.}~\bibnamefont
  {Ansorg}}, \bibinfo {author} {\bibfnamefont {B.}~\bibnamefont {Br\"ugmann}},
  \ and\ \bibinfo {author} {\bibfnamefont {W.}~\bibnamefont {Tichy}},\
  }\href@noop {} {\bibfield  {journal} {\bibinfo  {journal} {Phys. Rev.}\
  }\textbf {\bibinfo {volume} {D70}},\ \bibinfo {pages} {064011} (\bibinfo
  {year} {2004})},\ \Eprint {http://arxiv.org/abs/gr-qc/0404056}
  {gr-qc/0404056} \BibitemShut {NoStop}%
\bibitem [{\citenamefont {Healy}\ \emph {et~al.}(2017)\citenamefont {Healy},
  \citenamefont {Lousto}, \citenamefont {Nakano},\ and\ \citenamefont
  {Zlochower}}]{Healy:2017zqj}%
  \BibitemOpen
  \bibfield  {author} {\bibinfo {author} {\bibfnamefont {J.}~\bibnamefont
  {Healy}}, \bibinfo {author} {\bibfnamefont {C.~O.}\ \bibnamefont {Lousto}},
  \bibinfo {author} {\bibfnamefont {H.}~\bibnamefont {Nakano}}, \ and\ \bibinfo
  {author} {\bibfnamefont {Y.}~\bibnamefont {Zlochower}},\ }\href {\doibase
  10.1088/1361-6382/aa7929} {\bibfield  {journal} {\bibinfo  {journal} {Class.
  Quant. Grav.}\ }\textbf {\bibinfo {volume} {34}},\ \bibinfo {pages} {145011}
  (\bibinfo {year} {2017})},\ \Eprint {http://arxiv.org/abs/1702.00872}
  {arXiv:1702.00872 [gr-qc]} \BibitemShut {NoStop}%
\bibitem [{\citenamefont {Abbott}\ \emph
  {et~al.}(2020{\natexlab{a}})\citenamefont {Abbott} \emph
  {et~al.}}]{GW190521discovery}%
  \BibitemOpen
  \bibfield  {author} {\bibinfo {author} {\bibfnamefont {R.}~\bibnamefont
  {Abbott}} \emph {et~al.},\ }\href {\doibase 10.1103/PhysRevLett.125.101102}
  {\bibfield  {journal} {\bibinfo  {journal} {Phys. Rev. Lett.}\ }\textbf
  {\bibinfo {volume} {125}},\ \bibinfo {pages} {101102} (\bibinfo {year}
  {2020}{\natexlab{a}})}\BibitemShut {NoStop}%
\bibitem [{\citenamefont {Gayathri}\ \emph {et~al.}(2022)\citenamefont
  {Gayathri}, \citenamefont {Healy}, \citenamefont {Lange}, \citenamefont
  {O'Brien}, \citenamefont {Szczepanczyk}, \citenamefont {Bartos},
  \citenamefont {Campanelli}, \citenamefont {Klimenko}, \citenamefont
  {Lousto},\ and\ \citenamefont {O'Shaughnessy}}]{Gayathri:2020coq}%
  \BibitemOpen
  \bibfield  {author} {\bibinfo {author} {\bibfnamefont {V.}~\bibnamefont
  {Gayathri}}, \bibinfo {author} {\bibfnamefont {J.}~\bibnamefont {Healy}},
  \bibinfo {author} {\bibfnamefont {J.}~\bibnamefont {Lange}}, \bibinfo
  {author} {\bibfnamefont {B.}~\bibnamefont {O'Brien}}, \bibinfo {author}
  {\bibfnamefont {M.}~\bibnamefont {Szczepanczyk}}, \bibinfo {author}
  {\bibfnamefont {I.}~\bibnamefont {Bartos}}, \bibinfo {author} {\bibfnamefont
  {M.}~\bibnamefont {Campanelli}}, \bibinfo {author} {\bibfnamefont
  {S.}~\bibnamefont {Klimenko}}, \bibinfo {author} {\bibfnamefont {C.~O.}\
  \bibnamefont {Lousto}}, \ and\ \bibinfo {author} {\bibfnamefont
  {R.}~\bibnamefont {O'Shaughnessy}},\ }\href {\doibase
  10.1038/s41550-021-01568-w} {\bibfield  {journal} {\bibinfo  {journal}
  {Nature Astron.}\ }\textbf {\bibinfo {volume} {6}},\ \bibinfo {pages} {344}
  (\bibinfo {year} {2022})},\ \Eprint {http://arxiv.org/abs/2009.05461}
  {arXiv:2009.05461 [astro-ph.HE]} \BibitemShut {NoStop}%
\bibitem [{\citenamefont {Healy}\ and\ \citenamefont
  {Lousto}(2022)}]{Healy:2022wdn}%
  \BibitemOpen
  \bibfield  {author} {\bibinfo {author} {\bibfnamefont {J.}~\bibnamefont
  {Healy}}\ and\ \bibinfo {author} {\bibfnamefont {C.~O.}\ \bibnamefont
  {Lousto}},\ }\href {\doibase 10.1103/PhysRevD.105.124010} {\bibfield
  {journal} {\bibinfo  {journal} {Phys. Rev. D}\ }\textbf {\bibinfo {volume}
  {105}},\ \bibinfo {pages} {124010} (\bibinfo {year} {2022})},\ \Eprint
  {http://arxiv.org/abs/2202.00018} {arXiv:2202.00018 [gr-qc]} \BibitemShut
  {NoStop}%
\bibitem [{\citenamefont {Memmesheimer}\ \emph {et~al.}(2004)\citenamefont
  {Memmesheimer}, \citenamefont {Gopakumar},\ and\ \citenamefont
  {Schaefer}}]{Memmesheimer:2004cv}%
  \BibitemOpen
  \bibfield  {author} {\bibinfo {author} {\bibfnamefont {R.-M.}\ \bibnamefont
  {Memmesheimer}}, \bibinfo {author} {\bibfnamefont {A.}~\bibnamefont
  {Gopakumar}}, \ and\ \bibinfo {author} {\bibfnamefont {G.}~\bibnamefont
  {Schaefer}},\ }\href {\doibase 10.1103/PhysRevD.70.104011} {\bibfield
  {journal} {\bibinfo  {journal} {Phys. Rev. D}\ }\textbf {\bibinfo {volume}
  {70}},\ \bibinfo {pages} {104011} (\bibinfo {year} {2004})},\ \Eprint
  {http://arxiv.org/abs/gr-qc/0407049} {arXiv:gr-qc/0407049} \BibitemShut
  {NoStop}%
\bibitem [{\citenamefont {Tessmer}\ \emph {et~al.}(2010)\citenamefont
  {Tessmer}, \citenamefont {Hartung},\ and\ \citenamefont
  {Schafer}}]{Tessmer:2010hp}%
  \BibitemOpen
  \bibfield  {author} {\bibinfo {author} {\bibfnamefont {M.}~\bibnamefont
  {Tessmer}}, \bibinfo {author} {\bibfnamefont {J.}~\bibnamefont {Hartung}}, \
  and\ \bibinfo {author} {\bibfnamefont {G.}~\bibnamefont {Schafer}},\ }\href
  {\doibase 10.1088/0264-9381/27/16/165005} {\bibfield  {journal} {\bibinfo
  {journal} {Class. Quant. Grav.}\ }\textbf {\bibinfo {volume} {27}},\ \bibinfo
  {pages} {165005} (\bibinfo {year} {2010})},\ \Eprint
  {http://arxiv.org/abs/1003.2735} {arXiv:1003.2735 [gr-qc]} \BibitemShut
  {NoStop}%
\bibitem [{\citenamefont {Tessmer}\ \emph {et~al.}(2013)\citenamefont
  {Tessmer}, \citenamefont {Hartung},\ and\ \citenamefont
  {Schafer}}]{Tessmer:2012xr}%
  \BibitemOpen
  \bibfield  {author} {\bibinfo {author} {\bibfnamefont {M.}~\bibnamefont
  {Tessmer}}, \bibinfo {author} {\bibfnamefont {J.}~\bibnamefont {Hartung}}, \
  and\ \bibinfo {author} {\bibfnamefont {G.}~\bibnamefont {Schafer}},\ }\href
  {\doibase 10.1088/0264-9381/30/1/015007} {\bibfield  {journal} {\bibinfo
  {journal} {Class. Quant. Grav.}\ }\textbf {\bibinfo {volume} {30}},\ \bibinfo
  {pages} {015007} (\bibinfo {year} {2013})},\ \Eprint
  {http://arxiv.org/abs/1207.6961} {arXiv:1207.6961 [gr-qc]} \BibitemShut
  {NoStop}%
\bibitem [{\citenamefont {Damour}\ \emph {et~al.}(2008)\citenamefont {Damour},
  \citenamefont {Jaranowski},\ and\ \citenamefont {Schafer}}]{Damour:2007nc}%
  \BibitemOpen
  \bibfield  {author} {\bibinfo {author} {\bibfnamefont {T.}~\bibnamefont
  {Damour}}, \bibinfo {author} {\bibfnamefont {P.}~\bibnamefont {Jaranowski}},
  \ and\ \bibinfo {author} {\bibfnamefont {G.}~\bibnamefont {Schafer}},\ }\href
  {\doibase 10.1103/PhysRevD.77.064032} {\bibfield  {journal} {\bibinfo
  {journal} {Phys. Rev.}\ }\textbf {\bibinfo {volume} {D77}},\ \bibinfo {pages}
  {064032} (\bibinfo {year} {2008})},\ \Eprint {http://arxiv.org/abs/0711.1048}
  {arXiv:0711.1048 [gr-qc]} \BibitemShut {NoStop}%
\bibitem [{\citenamefont {Buonanno}\ \emph {et~al.}(2006)\citenamefont
  {Buonanno}, \citenamefont {Chen},\ and\ \citenamefont
  {Damour}}]{Buonanno:2005xu}%
  \BibitemOpen
  \bibfield  {author} {\bibinfo {author} {\bibfnamefont {A.}~\bibnamefont
  {Buonanno}}, \bibinfo {author} {\bibfnamefont {Y.}~\bibnamefont {Chen}}, \
  and\ \bibinfo {author} {\bibfnamefont {T.}~\bibnamefont {Damour}},\
  }\href@noop {} {\bibfield  {journal} {\bibinfo  {journal} {Phys. Rev.}\
  }\textbf {\bibinfo {volume} {D74}},\ \bibinfo {pages} {104005} (\bibinfo
  {year} {2006})},\ \Eprint {http://arxiv.org/abs/gr-qc/0508067}
  {gr-qc/0508067} \BibitemShut {NoStop}%
\bibitem [{\citenamefont {Buonanno}\ \emph {et~al.}(2011)\citenamefont
  {Buonanno}, \citenamefont {Kidder}, \citenamefont {Mroue}, \citenamefont
  {Pfeiffer},\ and\ \citenamefont {Taracchini}}]{Buonanno:2010yk}%
  \BibitemOpen
  \bibfield  {author} {\bibinfo {author} {\bibfnamefont {A.}~\bibnamefont
  {Buonanno}}, \bibinfo {author} {\bibfnamefont {L.~E.}\ \bibnamefont
  {Kidder}}, \bibinfo {author} {\bibfnamefont {A.~H.}\ \bibnamefont {Mroue}},
  \bibinfo {author} {\bibfnamefont {H.~P.}\ \bibnamefont {Pfeiffer}}, \ and\
  \bibinfo {author} {\bibfnamefont {A.}~\bibnamefont {Taracchini}},\ }\href
  {\doibase 10.1103/PhysRevD.83.104034} {\bibfield  {journal} {\bibinfo
  {journal} {Phys. Rev.}\ }\textbf {\bibinfo {volume} {D83}},\ \bibinfo {pages}
  {104034} (\bibinfo {year} {2011})},\ \Eprint {http://arxiv.org/abs/1012.1549}
  {arXiv:1012.1549 [gr-qc]} \BibitemShut {NoStop}%
\bibitem [{\citenamefont {Abbott}\ \emph
  {et~al.}(2020{\natexlab{b}})\citenamefont {Abbott} \emph
  {et~al.}}]{LIGOScientific:2020ufj}%
  \BibitemOpen
  \bibfield  {author} {\bibinfo {author} {\bibfnamefont {R.}~\bibnamefont
  {Abbott}} \emph {et~al.} (\bibinfo {collaboration} {LIGO Scientific,
  Virgo}),\ }\href {\doibase 10.3847/2041-8213/aba493} {\bibfield  {journal}
  {\bibinfo  {journal} {Astrophys. J. Lett.}\ }\textbf {\bibinfo {volume}
  {900}},\ \bibinfo {pages} {L13} (\bibinfo {year} {2020}{\natexlab{b}})},\
  \Eprint {http://arxiv.org/abs/2009.01190} {arXiv:2009.01190 [astro-ph.HE]}
  \BibitemShut {NoStop}%
\bibitem [{\citenamefont {Barrera}\ and\ \citenamefont
  {Bartos}(2022)}]{Barrera:2022yfj}%
  \BibitemOpen
  \bibfield  {author} {\bibinfo {author} {\bibfnamefont {O.}~\bibnamefont
  {Barrera}}\ and\ \bibinfo {author} {\bibfnamefont {I.}~\bibnamefont
  {Bartos}},\ }\href {\doibase 10.3847/2041-8213/ac5f47} {\bibfield  {journal}
  {\bibinfo  {journal} {Astrophys. J. Lett.}\ }\textbf {\bibinfo {volume}
  {929}},\ \bibinfo {pages} {L1} (\bibinfo {year} {2022})},\ \Eprint
  {http://arxiv.org/abs/2201.09943} {arXiv:2201.09943 [astro-ph.HE]}
  \BibitemShut {NoStop}%
\bibitem [{\citenamefont {Sch\"afer}\ and\ \citenamefont
  {Jaranowski}(2018)}]{Schafer:2018kuf}%
  \BibitemOpen
  \bibfield  {author} {\bibinfo {author} {\bibfnamefont {G.}~\bibnamefont
  {Sch\"afer}}\ and\ \bibinfo {author} {\bibfnamefont {P.}~\bibnamefont
  {Jaranowski}},\ }\href {\doibase 10.1007/s41114-018-0016-5} {\bibfield
  {journal} {\bibinfo  {journal} {Living Rev. Rel.}\ }\textbf {\bibinfo
  {volume} {21}},\ \bibinfo {pages} {7} (\bibinfo {year} {2018})},\ \Eprint
  {http://arxiv.org/abs/1805.07240} {arXiv:1805.07240 [gr-qc]} \BibitemShut
  {NoStop}%
\end{thebibliography}%

\end{document}